\documentclass[a4paper,12pt]{article}
\usepackage{latexsym,indentfirst}
\begin{document}
\pagestyle{plain}
\title{Predictions for the unitarity triangle angles in
a new parametrization}
\author{ S. Chaturvedi \thanks{scsp@uohyd.ernet.in}\\
\rm School of Physics, University of Hyderabad, \\
 Hyderabad 500 046 India\\
Virendra Gupta \thanks{virendra@aruna.mda.cinvestav.mx}\\
\rm Departamento de Fis\'ica Aplicada, CINVESTAV-Unidad M\'erida\\
 A.P. 73 Cordemex 97310 M\'erida, Yucatan, Mexico}
\date{\today}
\vskip-1cm
\maketitle
\begin{abstract}
 A new approach to the parametrization of the CKM matrix, $V$, is
 considered in which $V$ is written as a linear combination of
 the unit matrix $I$ and a non-diagonal matrix $U$ which causes
 intergenerational-mixing, that is  
$V=\cos\theta~ I+i\sin\theta~ U$. Such a $V$ depends on 3 real 
parameters including the parameter $\theta$. It is interesting 
that a value of $\theta=\pi/4$ is required to fit the available 
data on the CKM-matrix including CP-violation. Predictions of 
this fit for the angles $\alpha$, $\beta$ and $\gamma$ for the 
unitarity triangle corresponding to $V_{11}V^*_{13} + V_{21}
V^*_{23} +V_{31}V^*_{33} =0$, are given. For $\theta$=$\pi/4$, we 
obtain $\alpha=88.46^\circ$, $\beta=45.046^\circ$ and 
$\gamma=46.5^\circ$. These values are just about in agreement, 
within errors, with the present data. It is very interesting that 
the unitarity triangle is  expected to be approximately a  right-angled, 
isosceles triangle. Our prediction $\sin 2\beta=1$ is in excellent agreement 
with the value $0.99\pm 0.15 \pm 0.05$ reported by the Belle collaboration 
at the Lepton-Photon 2001 meeting.  
\end{abstract}
\newpage
\section{Introduction}
After the first explicit parametrization for three generations 
\cite{1}, many different parametrizations have been suggested \cite{2,3} for 
the Cabibbo-Kobayashi-Maskawa (CKM) matrix. Even today we do not have a
deep understanding of the observed mixing of the quark flavors in the
standard model.

Recently, a new approach to the parametrization of of the CKM-matrix, 
 $V$, was suggested motivated by the question whether $V$ consists of 
two parts. That is, a trivial part (taken to be the unit matrix $I$) for 
which the physical (or quark mass-eigenstate) basis and the the gauge 
basis are the same and a non-trivial part (represented by a non-
diagonal diagonal matrix $U$) for which the two bases are different and which 
causes quark flavor mixing. This possibility was explored in some 
detail in reference 4 where the CKM-matrix was taken to be given by 
the linear combination
\begin{eqnarray}
&& V(\theta)=\cos\theta~ I+i\sin\theta~U.
\end {eqnarray}
The value of $\theta$ determines the relative importance of the two 
parts. It also determines the magnitude of CP-violation in this 
approach. It was shown \cite{4} that the value of $\theta$ near $\pi/4$ 
gave a good fit to the data for the CKM-matrix given in year 1998 by the 
Particle Data Group \cite{3}.
      For $0<\theta<\pi/2$, it is clear that for $V$ to be unitary, $U$
  (independent of $\theta$) has to be hermitian and unitary. For three 
  generations, mathematically, such a $3\times 3$ matrix $U$ can depend on 
  at most 4 real parameters, namely, two moduli and two phases. In section 2, 
we give the explicit form of $U$ and $V$. It is shown that using the 
freedom of re-phasing transformations on $V$, one can eliminate the 
   two phases, so that, in effect, $U$ contains only 2 real parameters. 
Thus, $V$  depends on 3 parameters, one less than in the usual 
parametrizations.

In section 3, $V(\theta)$ is confronted with the data \cite{5} given
  in the year 2000 by the Particle Data Group. This update of the fits is
   necessary as the recent data differs from the earlier data of 1998. 
A satisfactory fit to the latest data is obtained and points strongly
    to a value of $\theta$ equal to $\pi/4$. 

In section 4, the predictions for the the angles of the triangle 
implied by the unitarity constraint
\begin{eqnarray}
       &&V_{11}V^*_{13} + V_{21}V^*_{23} +V_{31}V^*_{33} =0,
\end{eqnarray}
    are considered. This can be written as
   \begin{eqnarray}
              z_{1} + z_{2}  + z_{3}=0,
    \end{eqnarray}
  where the complex numbers $z_{i}$=$V_{i1}V^*_{i3}~; i=1,~2,~3$.
     In standard notation, the three angles of the triangle in terms
   of $z_{i}$ are $\alpha=arg(-z_{3}/ z_{1})$, $\beta=arg(- z_{2}/
  z_{3})$, and $\gamma=arg(-z_{1}/z_{2})$. Numerical results for these
  angles in our parametrization are compared with the available data. 

Finally, section 5, contains a brief summary with some 
speculative concluding remarks.

 \section{Parameters and form of $U$ and $V$}
    To determine the general form of $U$, one starts with a general
   hermitian matrix and then requires it to be unitary. For the case
   of three generations, it was shown earlier \cite{4} that such a matrix 
can be parametrized in terms of three complex numbers $a$, $b$ and $c$ which 
satisfy two constraints, one on their moduli and the other on their phases  
   $\phi_a$, $\phi_b$ and $\phi_c$. Explicitly,
   \begin{eqnarray}
        U=I -2\left(
       \begin{array}{ccc}
        |a|^2+|b|^2& b^*c& a^*c^*\\
         bc^*& |a|^2+|c|^2& ab^*\\
         ac& a^*b& |b|^2+|c|^2 \\
      \end{array}
       \right),
     \end{eqnarray}
  with the constraints
    \begin{eqnarray}
           |a|^2+|b|^2+|c|^2=1,
        \end{eqnarray}
     \begin{eqnarray}
    \phi_a - \phi_b + \phi_c = \pi/2.
     \end{eqnarray}
    Mathematically, this is the most general $U$ which is hermitian
    and unitary. It depends on four real parameters, two moduli and two 
phases. However, this $U$ is a part of the CKM-matrix $V$ for which one has
    the freedom to make re-phasing transformations without affecting
    its physical predictions. This freedom allows one to eliminate the
    phases in $U$ making it a real matrix depending on only two real
    positive parameters, the two moduli. In our ansatz, $V$ and its
    physical predictions depend on three real parameters, the angle $\theta$
    and the two moduli in $U$. This fact was not realised in reference 4.
         
       We consider the explicit phase transformation which eliminates
   the phases. Let $P(\lambda)$ denote the diagonal phase matrix, 
  diag $(e^{i\lambda_1},~e^{i\lambda_2},1)$. Then,
  \begin{eqnarray}
  &&V'(\theta)=P(\lambda)V(\theta) P^*(\lambda)=\cos\theta~ I+i\sin\theta~ U',
   \end{eqnarray}
    where the real matrix
    \begin{eqnarray}
       U'=I -2\left(
       \begin{array}{ccc}
        |a|^2+|b|^2&|bc|& -|ac|\\
         |bc|& |a|^2+|c|^2& |ab|\\
         -|ac|& |ab|& |b|^2+|c|^2 \\
      \end{array}\right),
    \end{eqnarray}
  is obtained by choosing the phases $\lambda_1=\phi_b-\pi/2$  and
    $\lambda_2=\phi_c-\pi/2$, while the phase constraint fixes $\phi_a$. 
It is interesting to note that the particular choice, $\phi_a=\phi_b
    =\phi_c=\pi/2$ (which respects the phase constraint) in Eq.$(6)$ will
    also give Eq.$(8)$. This was mistakenly referred to as a special case
    earlier \cite{4} as the above points were not realised there. The
    important point is that the physical predictions are independent of
   the   actual value of the individual phases of $a$, $b$ and $c$ as long as
   they satisfy the phase constraint.
     Since, under a phase transformation $|V_{ij}|=|V'_{ij}|$, it does 
    not matter whether $V$ or $V'$ is used to confront the data. In
     either case, $\theta$ and the two moduli will determine the matrix  
     elements of the CKM-matrix. Furthermore, the Jarlskog invariant \cite{6}, $J$, which gives CP-violation, is the same for $V$ or $V'$, 
namely
    \begin{eqnarray}
     J(V(\theta))=Im(V_{11}V_{22}V^*_{12}V^*_{21})
        =\cos\theta |V_{12}V_{13}V_{23}|=8\cos\theta \sin^3\theta |abc|^2. 
     \end{eqnarray}
     Note that $J$ is independent of the phases of $a$, $b$ and $c$.\\
    \indent To confront $V(\theta)$ with experiment we need to specify
      $\theta$. A simple and appealing choice is $\theta=\pi/4$ which
     gives equal weight to the two parts in $V$. In this case,
      \begin{eqnarray}
          &&V(\pi/4)=\frac{1}{\sqrt{2}}(I+iU).
     \end{eqnarray}

        In this case two experimental inputs e.g. $|V_{12}|$ and $|V_{23}|$
     are enough to determine all the the other $|V_{ij}|$. However, with
      3 inputs,$|V_{12}|$, $|V_{23}|$ and $|V_{13}|$ from the data will
    determine  all the $|V_{ij}|$ and also the value of $\sin\theta$. 
The numerical results for the 2 input $(\theta=\pi/4)$ and the 3
      input cases are given in the next section.
 \section { Fits to the recent data.}
  The experimentally determined CKM-matrix,$V_{EX}$, given by the
    Particle Data Group \cite{5}
    \begin{eqnarray}
    V_{EX}=\left(
    \begin{array}{ccc}
    0.9742-0.9757&0.219-0.226&0.002-0.005\\
    0.219-0.225&0.9734-0.9749&0.037-0.043\\
    0.004-0.014&0.035-0.043&0.9990-0.9993
    \end{array}\right).
   \end{eqnarray}
   The entries correspond to ranges for the moduli of the matrix
 elements. Since $U$ is hermitian, this implies that $|V_{ij}|$=$|V_{ji}|$   
 in our approach. It is clear that $|V_{12}|$=$|V_{21}|$ and
$|V_{23}|=|V_{32}|$ are satisfied for practically the whole range while
the equality $|V_{13}|=|V_{31}|$ is merely suggested by the data. Since
$|V_{13}|$ and $|V_{31}|$ are the most difficult to measure experimentally, it
is possible they may turn out to be equal. Note that for any unitary matrix
$V$, unitarity requires that either $|V_{ij}|=|V_{ji}|$ for all three
pairs or for none of them.
   To fit the data we convert the range for each modulus into a central
 value with errors, for example, $|V_{11}|=0.97495\pm 0.00075$. For each pair
of off-diagonal elements a common value was obtained by averaging the
 central values. For example, the common value 
$|V_{13}|=|V_{31}|=0.00625\pm 0.00325$ is obtained by taking the average
of the central values with errors of $|V_{13}|=0.0035\pm 0.0015$ and
$|V_{31}|=0.009\pm 0.005$. Similar procedure was used for the other two
pairs of off-diagonal matrix elements.
      
  a) Two parameter fit. We take the experimentally well determined
$|V_{12}|$ and $|V_{23}|$ as inputs. Given these, for general $\theta$, one
has
 \begin{eqnarray}
  |a|=|V_{23}|/(2\sin\theta |b|),
 \end{eqnarray}
 \begin{eqnarray}
  |c|=|V_{12}|/(2\sin\theta |b|).
  \end{eqnarray}
     The constraint, Eq.$(5)$, gives a quadratic for $|b|^2$ with the
solutions,
   \begin{eqnarray}
 |b|^2=\frac{1}{2} \left[  1\pm \sqrt{1-(|V_{12}|^2
+|V_{23}|^2)\csc^2\theta}   \right] .  
\end{eqnarray}
  Clearly, we need the positive solution since
  $|V_{12}|>|V_{23}|>|V_{13}|$.
  Further, for $|b|^2$ to be real, $\theta$ has to be greater than some
  minimum value. For the numerical input values $|V_{12}|=0.22225$ and
  $|V_{23}|=0.0395$, we need $\theta\geq 13.046^\circ$. For $\theta
  =\pi/4$,
  Eqs.(12-14) yield
 \begin{eqnarray}
  |a|=0.028303,\qquad |b|=0.986832,\qquad |c|=0.159251.
  \end{eqnarray}   
 The calculated values of $|V_{ij}|$ for $V(\pi/4)$ are given in Table I. 
These are to be compared with the central values of the
 experimental $|V_{ij}|$ given in the first column. The agreement is fairly
 good.

  The values of $J$ for $V_{EX}$ and $V(\pi/4)$ are also given in Table I. 
The value of $J(V_{EX})$ was calculated from the formula \cite{7}
 \begin{equation}
J^2=|V_{11}V_{22}V^*_{12}V^*_{21}|^2\
-\frac{1}{4} \left[1-|V_{11}|^2-|V_{22}|^2-|V_{12}|^2-
|V_{21}|^2+|V_{11}V_{22}|^2+|V_{12}V_{21}|^2\right]^2,\nonumber\\
 \end{equation} 
 using the central values of $|V_{ij}|~;~ i,j=1,2 $ since these four are the
 best measured. The value of $J(V(\pi/4))$ was calculated using Eq.$(9)$ and
 is about a factor 2 smaller than $J(V_{EX})$. The two values are in
 reasonable agreement considering the slight differences in the values of
 $|V_{ii}|~;~ i=1,2$ in the two cases and also because of the strong 
 numerical cancellation between the two terms on the rhs of Eq.$(16)$.\\
  \indent b)Three input fit. We now consider a three input fit which determines
 $\theta$ directly from the data. This can be done using the equation
  \begin{eqnarray}
  2\sin\theta=
 \frac{|V_{12}V_{23}|}{|V_{13}|}+\frac{|V_{12}V_{13}|}{|V_{23}|}
+\frac{|V_{13}V_{23}|}{|V_{12}|}.
 \end{eqnarray}
  With the numerical inputs $|V_{12}|=|V_{21}|=0.22225$,
 $|V_{23}|=|V_{32}|=0.0395$ and $|V_{13}|=|V_{31}|=0.00625$, we obtain 
$\sin\theta$ =0.720448 so that $\theta$ equal to $46.09^{\circ}$. This is
remarkably close to our earlier choice of  $\theta =\pi/4$. The values
obtained for the moduli are
  \begin{eqnarray}
   |a|=0.027765, \qquad |b|=0.987331,\qquad |c|=0.156223.
 \end{eqnarray}
   The calculated values of $|V_{ii}|;~i=1,2,3$ and $J$ are also given in 
Table I. These are practically same as the values for the two input
case. The reason is that the value obtained for $|V_{13}|$ for  $\theta
=\pi/4$ is very close to the input value used in $(17)$.
  The more recent data \cite{5} suggests more strongly than the earlier  
 data \cite{3} that  the value of $\theta=\pi/4$. In other words, the 
  diagonal part~($I$) and the non-diagonal part~($U$) in the CKM-matrix
  have equal weight.
 \section{Predictions for the angles $\alpha$, $\beta$ and $\gamma$} 
 To determine the angles of the unitarity triangle given by Eq.$(2)$, it is   
 convenient to denote the complex number
  \begin{eqnarray}
        -\frac{z_{1}}{z_{2}} =\rho +i\eta,
 \end{eqnarray}
    so that, using Eq.$(3)$,
  \begin{eqnarray}
    -\frac{z_{3}}{z_{2}} =(1-\rho) -i\eta.
  \end{eqnarray}
 The notation used here for the real and imaginary parts has been
 chosen so that it coincides with that used by Wolfenstein \cite{2}
 in his approximate parametrization of the CKM-matrix. This notation
  like that for the angles has become standard. However, the use of $\rho$
 and $\eta$ in Eqs.(19-20) is purely a matter of notation and the formulae
  such as Eqs.(21-22) below are valid for any exact parametrization of
 the CKM-matrix.
    From the definitions of the angles in section 1 and Eqs.(19-20)  
 it follows that
   \begin {eqnarray}
  \sin\alpha=\frac{\sin\beta}{\sqrt{\rho^2+\eta^2}}=
   \frac{\sin\gamma}{\sqrt{(1-\rho)^2+\eta^2}},
\end{eqnarray}
    and
 \begin{eqnarray}
         \tan\gamma = \eta/\rho .
   \end{eqnarray}
     Thus, the knowledge of $\rho$ and $\eta$ is sufficient to
   determine the three angles of the unitarity triangle.
      In general, $\rho$ and $\eta$ are independent parameters. However,
  in our parametrization, one expects a relation  between them \cite{8}
  since $V$ depends on 3 parameters only. Using the explicit form of
  $V$ in terms of $a$, $b$ and $c$ one obtains
  \begin{eqnarray}
      \rho= (1-2|c|^2)/2|b|^2   
           = -\frac{1}{2|b|^2}\frac{|V_{12}|^2-|V_{23}|^2}
       {|V_{12}|^2+|V_{23}|^2}+\frac{|V_{12}|^2}{|V_{12}|^2+|V_{23}|^2},
   \end{eqnarray}
    and
  \begin{eqnarray}
       \eta=\cot\theta/2|b|^2,
  \end{eqnarray}
    where $|b|^2$ is the positive solution given in Eq.$(14)$. The second  
    equality in Eq.$(23)$ requires the use of Eqs.$(12),~(13)$ and Eq.$(5)$.
       The relation between $\rho$ and $\eta$ is not simple. To derive it
   one notes that Eq.$(24)$ together with  Eq.$(14)$ can be used to solve for
 $|b|^2$ in terms of $\eta$ and $s= |V_{12}|^2 +|V_{23}|^2$. The result is
  \begin{eqnarray}
        \frac{s}{2|b|^2} =1 - \sqrt{1-s-\eta^2 s^2} .
  \end{eqnarray}   
  Substitution of this in Eq.$(23)$ gives $\rho$ in terms of
  $\eta$, $|V_{12}|$ and $|V_{23}|$. Numerical values of $|V_{13}|$, $\rho$, 
and $\eta$ for some representative values of $\theta$ are given in Table II. 
For input values $|V_{12}|$=0.22225
  and $|V_{23}|$=0.0395, as $\theta$ increases from its minimum value of
  13.046 to 90 degrees, $\rho$ increases from 0.03062 to 0.49386 and
  $\eta$ decreases from 4.31567 to 0. Algebraic expressions for the limits
  on $\rho$ and $\eta$ can be easily derived. Note, the variation with
  $\theta$ of $\eta$ is stronger than that of  $\rho$. The latter increases
  from about 0.47328 to 0.49386, while former decreases from 0.91532 to 0
  as $\theta$ goes from $\pi/6$ to $\pi/2$.
  For $\theta=\pi/4$, we obtain
         \begin{eqnarray}
     &&\eta=0.5134 \qquad \rho =0.4874.
        \end{eqnarray} 
 Using these values in Eqs.(21-22) gives
    \begin{eqnarray}
   &\alpha=88.46^\circ,  \beta=45.046^\circ,  \gamma=46.5^\circ.
  \end{eqnarray}
   So, the unitarity triangle is predicted to be approximately a
   right-angled isosceles triangle. The near isosceles nature of the
  triangle follows from the fact that in our parametrization $|V_{13}|=
  |V_{31}|$ and experimentally $|V_{11}|$ and $|V_{33}|$ are nearly
  equal.
   Since $\rho$ and $1-\rho$ are approximately equal over a wide
  of $\theta$ (approximately from 30 to 90 degrees), there is a simple
  mnemonic for the variation of the angles $\alpha$, $\beta$ and 
  $\gamma$ with $\theta$. A change in $\theta \to \theta + \delta$ implies
  approximately that $\beta \to \beta - \delta$, 
$\gamma \to \gamma - \delta$ while  $\alpha \to \alpha+
 2\delta$ to satisfy $\alpha +\beta+\gamma=\pi$. For example, for $\theta
 =30^\circ$ (that is, $\delta=-15^\circ$) one obtains from direct
calculations $\alpha=57.26^\circ,$ $\beta=60.08^\circ$ and
  $\gamma=62.66^\circ$, an approximate equilateral triangle!\\
   
A direct way to obtain the angles is to use Eq.$(21)$ in the form 
\begin{equation}
\frac{\sin\alpha}{|z_2|}=\frac{\sin\beta}{|z_1|}=\frac{\sin\gamma}{|z_3|}
\equiv \lambda > 0,
\end{equation}
where $\lambda$ is a positive real number. Since 
$\sin\alpha=\sin(\beta+\gamma)$, this gives 
\begin{equation}
4|z_1z_2z_3|^2~\lambda^2 = (|z_1|^2+|z_2|^2+|z_3|^2)^2
 -2(|z_1|^4+|z_2|^4+|z_3|^4).
\end{equation} 
 
Since $|z_1| =|V_{11}||V_{13}|$ etc. are known one can determine $\lambda$ 
and hence the angles. The results are tabulated in Table III. Using the values 
of $|V_{ij}|$ for the three cases in Table I one obtains  nearly the same 
values, for $\lambda$ and the angles. Note that the values of the angles,   
for the 2-input case when $\theta=\pi/4$, are very close to those in Eq.$(27)$
as they should be.

The values of the angles in Eq.$(27)$ imply that $\sin2\alpha$=0.054
  and $\sin2\beta$=1. The former and the value of $\gamma$ in Eq.$(27)$
  are compatible with a recent theoretical analysis \cite{9}, but which 
  obtains $0.49 < \sin2\beta <0.94$. The data has a large variation.
  Our result for $\sin2\beta$ is compatible within errors with the
 values $0.79^{+0.41}_{- 0.44}$  and  $0.84^{+0.82}_{-1.04}\pm0.16$
obtained by the CDF and ALEPH collaborations \cite{10}. The values
\cite{11} quoted by BaBar $(0.34\pm0.20\pm0.05)$ and Belle $(0.58^{+0.32
+0.09}_{-0.34- 0.10})$ are lower, especially that of BaBar. However, the 
values reported at the Lepton-Photon 2001 \cite{12, 13} are much 
larger. Babar \cite{12} and Belle \cite{13} give the values 
$0.59\pm 0.14 \pm 0.05$ and $0.99 \pm 0.14 \pm 0.06$ respectively. The 
latter agrees with our prediction.  Experiments with high statistics are in 
progress at both Belle and BaBar and hopefully more definitive values for 
the angles will be available in a year or so.
\section{Summary and concluding remarks}
  The ansatz $V(\theta)=\cos\theta~ I+i\sin\theta ~ U $, considered here, 
was motivated by the question whether the CKM-matrix is a linear
  combination of a trivial part ($I$) and a non-trivial part ($U$). The  
   matrix $U$ depends on 2 real parameters and so that the CKM-matrix   
   $V$ depends on 3 real parameters including $\theta$ which plays a
   double role. It determines the relative importance of the generation
  mixing term ($U$) relative to the generation diagonal term ($I$). It also 
  determines the magnitude of CP-violation. The fits to the presently
   available data \cite{5} for the CKM-matrix require that  
  $\theta=\pi/4$, that is $V(\pi/4)=(I+iU)/\sqrt{2}$, implying equal
 importance of the two parts. This fit predicts that the unitarity
triangle is approximately a right-angled isosceles triangle 
 (see section 4). These results are compatible with present data wthin
  allowed errors. Experiments, in the offing, will soon decide the validity
  of the approach used here for parametrizing the CKM-matrix.\\
   \indent  The  ansatz  considered  here  has  been   proposed   
\cite{14} recently   for   the \\
 parametrization of the neutrino mixing
matrix $V_\nu$, the subscript  `$\nu$' will denote the corresponding quantities
for the neutrino sector. Remarkably, one finds that
$V_\nu=(I+iU_\nu)/\sqrt{2}$ can explain the neutrino data with $U_\nu$ 
depending on only one small parameter. In this case, the atmospheric neutrino
data \cite{15}, which requires maximal $\nu_\mu$ and $\nu_\tau$ mixing, 
forces the value of
$\theta_\nu$ to be equal to $\pi/4$. It is extremely interesting that in both 
the lepton and quark sectors the relative weight of the diagonal part
  and the non-diagonal pieces in the mixing matrix are the same! Even
  though $V$ and $V_\nu$ are very different, one may speculate that an  
  underlying quark-lepton symmetry seems to be suggested, in our approach,
  by the equality $\theta =\theta_\nu = \pi/4$ and furthermore, this
  equality may emerge naturally in a grand unification model.\\
\vskip1.0cm  
\indent \textbf{Acknowledgements}. Major part of the  work in this paper
was done while one of us (VG) was visiting  Stanford
 Linear Accelerator Center, California, USA and  the Department   
  of Physics, National Taiwan University, Taipei. He wishes to 
acknowledge with gratitude the hospitality of the
  latter institution and also the help he received there 
from Mr. Ching-Cheng Hsu in the preparation of this manuscript. The work 
was completed during his visit to the University of Hyderabad, Hyderabad, 
India, during August 2001.

\newpage
\begin{tabular}{|l|l|l|r|}  \hline \hline
QUANTITY & EXPERIMENT & \multicolumn{2}{c|}{THEORY} \\
 \hline
$|V_{12}|=|V_{21}|$ & 0.22225 $\pm 0.00325$ & INPUT & INPUT \\ \hline
$|V_{23}|=|V_{32}|$ & 0.0395 $\pm 0.0035$ & INPUT & INPUT\\ \hline
$|V_{13}|=|V_{31}|$ & 0.00625$\pm 0.00325$ & 0.006374369 & INPUT \\ \hline
$ |V_{11}| $ & 0.97495$\pm 0.00075$ & 0.97496887 & 0.97496968 \\ \hline
$ |V_{22}| $ & 0.97415$\pm 0.00075$ & 0.97418925 & 0.97418925 \\ \hline
$ |V_{33}| $ & 0.99915$\pm 0.00015$ & 0.99919924 & 0.99920002
\\ \hline
J$\times 10^{5}$ & 6.7 & 3.96 & 3.8 \\ \hline
\end{tabular}
\vskip0.5cm
Table I  Numerical values of the moduli of the matrix elements of the
CKM-matrix $V$. Experimental quantities are average values obtained from 
$V_{EX}$ in Eq.(11). The errors reflect the range of the values of 
$|V_{ij}|$ as explained in the text. Column 3 gives the results for
the 2 input fit with $\ \theta =\pi /4.$ Column 4 gives the results for the
3 input fit ( see section 3).
\vskip1.0cm

\begin{tabular}{|c|c|c|c|}  \hline \hline
$\theta$ & $|V_{13}|$ & $\rho$ &$\eta$ \\ \hline
$\theta_{min}=13.046^\circ$ &0.0388906 &0.03062 &4.31567\\ \hline
$30^\circ$ & 0.00927858 & 0.473282 & 0.915321 \\ \hline
$40^\circ$  &0.00705337 & 0.484561 & 0.615477\\ \hline
$45^\circ$ & 0.00637437 & 0.48739 & 0.513432 \\ \hline
$50^\circ$ & 0.0058601 & 0.489343 & 0.429076 \\ \hline
$60^\circ$ & 0.00515763 & 0.491745 & 0.293752 \\ \hline
$90^\circ$ & 0.00444683 & 0.493863 & 0.0 \\ \hline
\end{tabular}
\vskip0.5cm
Table II Numerical values of $|V_{13}|,~\rho$ and $\eta$ as a function of 
$\theta$. While $\rho$ and $\eta$ are calculated using Eqs. $(24), (25)$, 
$|V_{13}|$ is calculated using $|V_{13}|= (|V_{12}||V_{23}|/s)(\sin \theta 
-\sqrt{\sin^2\theta-s})$.
\vskip1.0cm

\begin{tabular}{|c|c|c|c|}  \hline \hline
QUANTITY & EXPERIMENT & \multicolumn{2}{c|}{THEORY} \\
\cline{3-4}
         &            &2-inputs & 3-inputs\\ \hline
$\lambda$ & 113.9017 & 113.8689 & 113.9012 \\ \hline
$\alpha$ & 89.317 & 88.46 & 89.29\\ \hline
$\beta $ & 43.952 & 45.046 & 43.953 \\ \hline
$ \gamma $& 45.358 & 46.49 & 45.342 \\ \hline
\end{tabular}
\vskip0.5cm
Table III  Values of the angles obtained directly from the general relation 
between the sides and angles satisfied by any triangle given in Eq.$(28)$.

\end{document}